\documentclass[twocolumn,showpacs,preprintnumbers,amsmath,amssymb,prl]{revtex4}

\usepackage{graphicx}% Include figure files
\usepackage{dcolumn}% Align table columns on decimal point
\usepackage{bm}% bold math
\usepackage[normalem]{ulem}
\newcommand{\be}{\begin{equation}}
\newcommand{\ee}{\end{equation}}
\newcommand{\bea}{\begin{eqnarray}}
\newcommand{\eea}{\end{eqnarray}}
\newcommand{\beaa}{\begin{eqnarray*}}
\newcommand{\eeaa}{\end{eqnarray*}}
\newcommand{\ben}{\begin{enumerate}}
\newcommand{\een}{\end{enumerate}}
\newcommand{\bi}{\begin{itemize}}
\newcommand{\ei}{\end{itemize}}
\newcommand{\lip}{\langle}
\newcommand{\rip}{\rangle}
\newcommand{\uu}{\underline}

\newcommand{\df}{{\rm d}}

\begin{document}

\preprint{}

\title{Sudden Death of Entanglement induced by Polarization Mode Dispersion}
\author{Cristian Antonelli,$^1$ Mark Shtaif,$^2$ and Misha Brodsky$^3$}%
%\email{cristian.antonelli@univaq.it}
\affiliation{%
$^1$Dipartimento di Ingegneria Elettrica e dell'Informazione and CNISM, Universit\`a dell'Aquila, 67040 L'Aquila,
Italy
}%
%\author{Mark Shtaif}
\affiliation{%
$^2$School of Electrical Engineering, Tel-Aviv University, Tel-Aviv 69978, Israel
}%
%\author{Misha Brodsky}
\affiliation{%
$^3$AT\&T Labs, 200 Laurel Ave. S., Middletown, NJ 07748 USA
}%

\date{\today}

\begin{abstract}
We study the decoherence of polarization-entangled photon pairs subject to the effects of polarization mode dispersion, the chief polarization decoherence mechanism in optical fibers. We show that fiber propagation reveals an intriguing interplay between the concepts of entanglement sudden death, decoherence-free sub-spaces and non-locality. We define the boundaries in which entanglement-based quantum communications protocols relying on fiber propagation can be applied.
\end{abstract}



\maketitle

%%%%%%%%%%%%%%%%%%%%%%

Entanglement between particles is a fundamental feature of quantum physics. Just as fundamental is the phenomenon of decoherence that takes place when the entangled quantum system interacts with the environment. One of the most intriguing recent discoveries related to decoherence is the phenomenon of entanglement sudden death (ESD) \cite{Sd_Eberly,Sd_Yu}. It manifests itself in an abrupt disappearance of entanglement once the interaction with the environment reaches a certain threshold. \cite{Sd_Almeida,Sd_Kimble,Ann}. Beyond the interest that it attracts as a fundamental physical phenomenon, decoherence plays a central role in quantum communications. The security of recent quantum key distribution protocols explicitly relies on the nonlocal properties of entanglement, quantified in terms of the violation of a Bell-type inequality \cite{Ekert,Acin_NJP06,Acin_PRL07}. Therefore establishing the relation between the violation of non-locality and ESD, which is very interesting from a standpoint of basic physics, has a potentially large impact on the new area of quantum communications.

A configuration that provides an excellent platform for the controlled study of decoherence is that of polarization entangled photon-pairs, distributed over optical fibers. In this scheme the main source of decoherence is the residual optical birefringence randomly accumulating along the fiber. While an alternative entanglement scheme, insensitive to birefringence, has been proposed \cite{Thew,Tanzilli}, the ease with which light polarization can be manipulated using standard instrumentation leaves polarization-entanglement the configuration of choice in many situations \cite{PoppeAndHubel}. Moreover, numerous sources of polarization entangled photons suitable for use with standard fibers have recently become available \cite{Shawn}. Hence, understanding the relation between non-locality and ESD as well as the ultimate limits imposed by fiber birefringence on the distribution of polarization-entangled photons in fibers is a problem of utmost importance.

The fact that optical birefringence is a major polarization decoherence mechanism has been known for a while. Indeed birefringent crystals have been used extensively for the creation and manipulation of special quantum states, such as the MEMS \cite{BarbieriPRL04,PetersPRL04} or Werner states \cite{BarbieriPRL04,WhitePRA01}. Similarly, birefringent crystals have also been used for the controlled demonstration of decoherence-free subspaces \cite{Kwiat_science,Altepeter_PRL}. Yet, the arbitrary birefringence characterizing fiber-optic transmission, produces a previously unobserved combination of physical effects.

The accumulation of randomly varying birefringence in fibers leads to a phenomenon known by the name of polarization mode dispersion (PMD) \cite{Gordon&Kogelnik}. Since the analysis of the general case of PMD is quite cumbersome, we limit ourselves to the simplest regime of operation in which the optical bandwidth of the photons is small in comparison with the bandwidth over which PMD decorrelates \cite{PMD_bandwidth}. In this regime, without loosing the essence of the problem, the overall effect of PMD resembles that of pure birefringence in the sense that it causes an incident pulse to split into two orthogonally polarized components delayed relative to each other \cite{Gordon&Kogelnik}. The polarization states of these two components are known as the principal states of polarization (PSP) and the delay between them is called the differential group delay (DGD).

In contrast to the controlled environment of \cite{BarbieriPRL04}-\cite{Altepeter_PRL}, both the PSP and the DGD of real fibers vary stochastically in time. Since typical time constants characterizing the decorrelation of PMD in optical fibers are as long as hours, days and sometimes months \cite{Misha}, PMD evolution can be considered adiabatic in the context of quantum communications protocols. Thus the density matrix describing the quantum state needs to be evaluated as a function of the arbitrary values of the instantaneous PMD. As a consequence, the parameters of interest obtained from the so evaluated density matrix are also PMD dependent. The temporal statistics for those parameters could be in principle determined by application of the proper PMD statistics. An approach, in which the randomness of PMD is accounted for in the density matrix itself \cite{Poon}, implicitly assumes ultrafast PMD dynamics and leads to fundamentally different results. This previously unstudied reality, produces important consequences to the dynamics of decoherence between polarization entangled photons.

In this letter we formulate a quantitative approach to studying PMD-induced disentanglement. We consider the evolution of an arbitrary two-photon state maximally entangled in polarization as each photon propagates through a fiber with PMD. Our studies demonstrate that the unequal and increasing differential delays in both arms always lead to \textit{entanglement sudden death} \cite{Sd_Eberly,Sd_Yu} for all but two special PSPs orientations. That is, our channel causes an abrupt drop to zero in concurrence while a single photon subjected to the same environment depolarizes asymptotically. Contrary to that, when \textit{both} delays are sufficiently close in value there exists a range of PSP orientations for which concurrence does not vanish. This is related to the existence of decoherence free subspaces and offers an opportunity for non-local PMD compensation. Finally, when only one photon experiences PMD, the concurrence decays gradually for every PSP orientation. Besides concurrence, we calculate the $S$ parameter of the Clauser-Horne-Shimony-Holt Bell's inequality. Comparing the loss of entanglement (concurrence $C=0$) and violation of locality ($S>2$) we discover an intriguing empirical relation between them.

We assume a source in which a pair of polarization entangled photons is generated either via spontaneous parametric down conversion \cite{SPDC}, or by using four--wave mixing \cite{Chi3_Kumar,Chi3_Takesue}. In either case, the quantum state of the generated photon-pair can be expressed as
\bea|\psi\rip=\int\frac{\df \omega}{2\pi} \tilde f(\omega)|\omega,-\omega\rip \otimes\frac{|\uu u_A,\uu u_B\rip+e^{i\alpha}|\uu u'_A,\uu u'_B\rip}{\sqrt{2}}, \label{Tensor1}\eea
where the left-hand side of the tensor product represents the frequency waveform, whereas the right-hand side represents polarization modes. The frequency variable $\omega$ denotes the offset from the central frequency, which is equal to the pump-frequency in sources relying on four--wave mixing and to half the pump frequency in sources based on spontaneous parametric down conversion. The function $\tilde f(\omega)$ represents the effect of phase-matching as well as the possible effects of filters, as we shall see below. Notice that normalization of the state $|\psi\rip$ implies that $\int\df\omega|\tilde f(\omega)|^2=2\pi T^{-1}$ with $T$ being the integration time of the detectors used in the set-up. In what follows, the state (\ref{Tensor1}) will be referred to as the \emph{input state} of the system, which consists of the two optical paths that lead the entangled photons from the source of entanglement towards its users, conventionally referred to as Alice and Bob. The terms $\uu u_{A,B}$ and $\uu u'_{A,B}$ are the Jones vectors that correspond to the excited polarization states of Alice's and Bob's photons, respectively. We use primes to denote orthogonality in polarization space, so that $\uu u_{A,B}\cdot\uu u'_{A,B}=0$. The phase factor $\alpha$ is introduced for consistency with the experimental generation of polarization entangled photon pairs \cite{Chi3_Kumar}. Notice that the frequency dependent part in (\ref{Tensor1}) can be re-expressed as
\bea\int\frac{\df \omega}{2\pi} \tilde f(\omega)|\omega,-\omega\rip=\iint\df t_A\df t_B f(t_A-t_B)|t_A,t_B\rip\label{psi_t}\eea
with $f(\tau)$ being the inverse Fourier transform of $\tilde f(\omega)$. As can be deduced from the form of Eq. (\ref{psi_t}), $|f(\tau)|^2$ is proportional to the probability density function that Bob's photon precedes Alice's photon (or vice versa) by $\tau$. For brevity, we will denote the polarization dependent part in the tensor product  (\ref{Tensor1}) by $|\psi_p\rip=\left[|\uu u_A,\uu u_B\rip+e^{i\alpha}|\uu u'_A,\uu u'_B\rip\right]/\sqrt{2}$, whereas the expression in Eq. (\ref{psi_t}) will be shortly denoted as $|f(t_A-t_B)\rip$. The overall state is then expressed as $|\psi\rip= |f(t_A-t_B)\rip\otimes|\psi_p\rip$.

Let us now represent the state $|\psi\rip$ in terms of the principal states of the PMD in the two arms. We denote by $\{\uu s_A,\uu s'_A\}$ and $\{\uu s_B,\uu s'_B\}$ the pairs of Jones vectors that correspond to the PSP along the paths of photons A and B, respectively. We now represent $|\psi_p\rip$ in the basis of the PSP modes as follows
\bea |\psi_p\rip &=&  \eta_1 \left( |\uu s_A, \uu s_B \rip  + e^{i \tilde \alpha_1}  | \uu s_A', \uu s_B' \rip \right) / \sqrt 2 \nonumber \\ &+& \eta_2 \left( | \uu s_A, \uu s_B' \rip - e^{i \tilde \alpha_2} | \uu s_A', \uu s_B \rip \right) / \sqrt 2 ,\label{Launch}\eea
where the coefficients $\eta_1$ and $\eta_2$ are given by
\bea \eta_1 = \left( \uu s_A\cdot\uu u_A  \right) \left( \uu s_B\cdot\uu u_B  \right) + e^{i\alpha} \left( \uu s_A\cdot\uu u'_A  \right) \left( \uu s_B\cdot\uu u'_B  \right)&&\label{1014} \\
\eta_2 =  \left( \uu s_A\cdot\uu u_A  \right) \left( \uu s_B'\cdot\uu u_B  \right) + e^{i\alpha} \left( \uu s_A\cdot\uu u'_A  \right) \left( \uu s_B'\cdot\uu u'_B  \right)  \label{1015}&&\eea
and where $\tilde \alpha_i$ is defined through the relation  $\eta_i=|\eta_i|\exp\big(i(\alpha-\tilde\alpha_i)/2\big)$. Also, note that, as is implied by state normalization, $|\eta_1|^2+|\eta_2|^2=1$. The quantity $|\eta_1|^2$ is related to the alignment between the input two-photon state (\ref{Tensor1}) and the PSP. Thus, for example, in the case of $\uu u_A=\uu s_A$ and $\uu u_B=\uu s_B$, the value of $|\eta_1|^2$ is unity. In the presence of PMD the arrival time of the $A$-photon is delayed by $\tau_A/2$ in the $\uu s_A$ polarization and advanced by the same amount in the $\uu s'_A$ polarization, and the $B$-photon undergoes a similar process. Therefore, the output state, i.e. the two-photon state after propagating through media with PMD, can be expressed as
\bea |\psi_{\mathrm{out}}\rip &=&  \frac{\eta_1}{\sqrt 2} | f(t_A -t_B-\frac{\tau_A -\tau_B}{2})\rip\otimes |\uu s_A, \uu s_B \rip \nonumber \\ &+& \frac{\eta_2}{\sqrt 2} | f(t_A -t_B-\frac{\tau_A +\tau_B}{2} )\rip\otimes |\uu s_A, \uu s_B' \rip\nonumber\\ &-& \frac{\eta_2^* e^{i\tilde\alpha_2}}{\sqrt 2} | f(t_A -t_B+\frac{\tau_A +\tau_B}{2} )\rip\otimes|\uu s_A', \uu s_B \rip\nonumber\\ &+& \frac{\eta_1^* e^{i \tilde\alpha_1}}{\sqrt 2}  | f(t_A -t_B+\frac{\tau_A -\tau_B}{2})\rip\otimes|\uu s_A', \uu s_B' \rip.\,\,\,\,\,\,\,
\label{psi_out}\eea
The density matrix that characterizes the detected field is given by
$ \rho=\int\df t'_A\df t'_B \lip t'_A,t'_B|\psi_{\mathrm{out}}\rip\lip\psi_{\mathrm{out}}|t'_A,t'_B\rip$,
where tracing over the time modes is performed in order to account for the fact that the photo-detection process is not sensitive to the photon's time of arrival (within the detector's integration window). The elements of the resulting density matrix, establishing the correspondence $(\uu s_A, \uu s_B) \leftrightarrow 1$, $(\uu s_A', \uu s_B) \leftrightarrow 2$, $(\uu s_A, \uu s_B') \leftrightarrow 3$, and $(\uu s_A', \uu s_B') \leftrightarrow 4$ are then given by
\bea \rho_{11} &=& \rho_{44} = \left| \eta_1 \right|^2/2  \nonumber \\ \rho_{22} &=& \rho_{33} =  |\eta_2|^2/2 \nonumber \\
\rho_{31} &=& - \rho_{42} =  \eta_1^* \eta_2 R_f\left(  \tau_B \right)/2   \nonumber \\
\rho_{21} &=& - \rho_{43} = -  \eta_1^* \eta_2^* e^{i \alpha} R_f\left(  \tau_A \right)/2  \nonumber
\\
\rho_{41} &=&  \left(\eta_1^*\right)^2 e^{i \alpha} R_f\left( \tau_A-\tau_B \right) /2 \nonumber \\
\rho_{23} &=& - \left(\eta_2^*\right)^2 e^{i \alpha} R_f\left( \tau_A+\tau_B \right) /2,
\label{1012} \eea
where $R_f(\tau)=T\int \df t f^*(t)f(t+\tau)$ is the autocorrelation function of $f(t)$, normalized such that $R_f(0)=1$.
The function $\tilde f(\omega)$ which defines the frequency contents of the two generated photons accounts for the phase matching spectrum, as well as for filtering applied to the two generated photons. In this case $\tilde f(\omega)=\tilde f_{\textrm{pm}}(\omega) H_A(\omega)H_B(-\omega)$, where $H_A(\omega)$ and $H_B(\omega)$ denote the transfer functions of Alice's and Bob's filters, respectively, and where $\tilde f_{\textrm{pm}}(\omega)$ represents the phase-matching spectrum. In most applications, the filters are much narrower than the phase matching spectrum, in which case $\tilde f_{\textrm{pm}}(\omega)$ can be replaced by a constant, such that $\tilde f(\omega)\propto H_A(\omega)H_B(-\omega)$. Notice that the effect of PMD scales with the width of the autocorrelation function $R_f(\tau)$, which is in turn determined by the overlap bandwidth of Alice's and Bob's filters, namely by the width of $|H_A(\omega)H_B(-\omega)|^2$. For illustrative purposes, in all the numerical examples considered in what follows, we will assume that $|H_A(\omega) |^2$ and $|H_B(\omega)|^2$ are Gaussian functions of root mean square bandwidth $B$ and central frequencies $\omega_A$ and $\omega_B=-\omega_A$ respectively.

\begin{figure}
\begin{centering}
\includegraphics[width=.95\columnwidth]{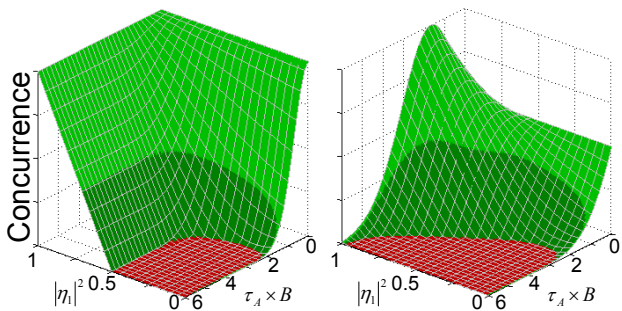}
\caption{Concurrence versus $\tau_A$ and $|\eta_1|^2$. In (a) $\tau_B = \tau_A$, whereas in (b) $\tau_B \simeq 1.7B^{-1}$. Light-green: $S>2$. Dark-green: $S<2$. Red: $C=0$ and $S<2$.} \label{Figure1}
\end{centering}
\end{figure}

We now turn to the characterization of the degree of entanglement of the PMD-affected two-photon state. In the presence of PMD, tracing over the time of arrival puts the system in a partially mixed state. The extent of this process can be quantified with the help of several proposed entanglement metrics \cite{Sergienko,HorodeckiRev,Wootters}. We choose to calculate purity, concurrence \cite{Wootters} and Bell's $S$ parameter, thus assessing the largest possible violation of Bell's inequality in the CHSH definition \cite{CHSH}. Because the individual photon states are maximally mixed, the two-photon density matrix can be reduced to a Bell-diagonal form by a proper change of basis \cite{Supplement}. This enables a fully analytical evaluation of $C$ and $S$. Note that a Bell-diagonal matrix is defined by three real parameters only. In the case of PMD they are $\tau_A$, $\tau_B$ and $|\eta_1|^2$ and the functional dependence of $C$ and $S$ on them is given in \cite{Supplement}.

In the simplest case of PMD present in only one of the two fibers, as described, for example, by $\tau_B=0$, the concurrence is given by $C = |R_f(\tau_A)|$; in this case the concurrence is independent of the PSP orientation and can only decay asymptotically with $\tau_A$. Correspondingly, the $S$ parameter acquires the maximum value compatible with such concurrence, that is $S = 2\sqrt{1+C^2}$ \cite{Verstraete}, indicating unconditional violation of Bell's inequality. Note that the two cases $|\eta_1|=1$ and $|\eta_1|=0$, where concurrence simplifies to $C = |R_f(\tau_A-\tau_B)|$ and $C = |R_f(\tau_A +\tau_B)|$ respectively, are equivalent to that of single-arm PMD, with corresponding nonzero DGDs equal to $\tau_A-\tau_B$ and $\tau_A+\tau_B$. Remarkably, for $\tau_A = \tau_B$ and $|\eta_1|=1$, the concurrence is unity, regardless of the DGD magnitude. This result is quite interesting and it is directly related to the concept of decoherence-free subspaces \cite{Kwiat_science,Altepeter_PRL}. In this situation the output state, when expressed in the basis of the PSP, is a superposition of a state in which both photons are delayed, with a state in which they are both advanced (see Eq. (\ref{Launch})), such that they reach the detectors simultaneously. Since in this state, knowledge of the photon's times of arrival discloses no information on their polarization states, tracing out time involves no loss of information. Decoherence-free subspaces would not be allowed if PMD dynamics were fast on the scale of measurements, as assumed in \cite{Poon}.

The dependence of the two-photon state decoherence on the PMD parameters in the general case is more cumbersome, as it is governed by the two DGD values $\tau_A$ and $\tau_B$ and by the PSP orientation, accounted for by $|\eta_1|$. For illustrative purposes, we plot in Fig. \ref{Figure1} the concurrence as a function of $\tau_A$ (normalized to $B^{-1}$) and $|\eta_1|^2$ for two different settings of $\tau_B$, so as to describe PMD effects for the most relevant realizations of PMD parameters. In Fig. \ref{Figure1}a we plot the concurrence for the case of identical DGDs, $\tau_B = \tau_A$, whereas in Fig. \ref{Figure1}b the value of $\tau_B$ is fixed and equal to $1.7B^{-1}$. The range of values for which entanglement disappears entirely (i.e. $C=0$) is emphasized in red color, whereas the green colored regions of the surface correspond to settings in which entanglement exists (i.e. $C>0$). In the light green area $S>2$, whereas in the dark green area $S\le2$. In the case where the DGD values are equal (Fig. 1a), the concurrence approaches unity when $|\eta_1|^2\to 1$. That is because in this situation the input state is given by the first term in Eq. (\ref{Launch}), and it is not affected by the loss of time of arrival information. In the opposite limit, when $|\eta_1| = 0$, the concurrence is given by $C = |R_f(2\tau_A)|$ and reduces towards zero asymptotically for large DGD values. Figure \ref{Figure1}b also illustrates that concurrence can be unity only when the DGD values in the two arms are equal.

The most interesting feature in Fig. \ref{Figure1} is the abrupt transition of the concurrence to zero when either $\tau_A$, or $|\eta_1|^2$ are varied continuously. The abrupt decay of concurrence is contrasted with the asymptotic decoherence that would be experienced by a single photon if it evolved in the same environment. Indeed, the purity of a polarized single-photon pulse characterized by a Jones vector $\uu u$ and a time-mode distribution $g(t)$, transmitted through a fiber with DGD $\tau$ and PSP $\uu s$, is given by $p = 1 - 2|\uu u \cdot \uu s|^2 \left( 1 - |\uu u \cdot \uu s|^2 \right)\left[ 1 - |R_g(\tau)|^2 \right]$, with $R_g(\tau)$ the autocorrelation function of $g(t)$.  This is a manifestation of the entanglement sudden-death (ESD) that has been previously reported in other physical systems \cite{Sd_Eberly,Sd_Yu,Sd_Almeida,Sd_Kimble,Ann}.

\begin{figure}
\begin{centering}
\includegraphics[width=.75\columnwidth]{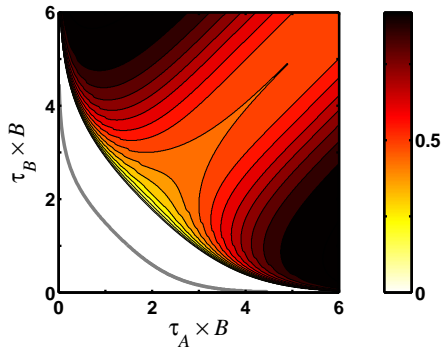}
\caption{ESD probability (see text) versus $\tau_A$ and $\tau_B$ normalized to $B^{-1}$. The grey line is the boundary $S=2$. Below and to the left of the grey line $S>2$ for all values of $|\eta_1|^2$, whereas to its right $S<2$ for some range of $|\eta_1|^2$.} \label{Map}
\end{centering}
\end{figure}

In general, PSP orientation in optical fibers varies faster than the DGD, resulting in a uniform distribution of the parameter $|\eta_1|^2$ between 0 and 1. For each given combination of $\tau_A$ and $\tau_B$, we evaluate the fraction of the interval between 0 and 1 in which ESD occurs. This quantity, which can be interpreted as the probability of ESD (conditioned on the DGD values) is illustrated in Fig. 2. Note the existence of a completely white region that shows the range of DGD values in which entanglement does not disappear for any value of $|\eta_1|^2$. In contrast to that, the darker tones mark areas for which ESD occurs for some range of $|\eta_1|^2$. The color progressively turns dark for highly differing DGD values. On the other hand, when increasing DGDs remain nearly equal, the probability of ESD is about 0.5 \cite{Supplement}.

Many applications involving entanglement rely directly on the violation of Bell's inequality \cite{Acin_NJP06,Acin_PRL07} and, therefore, we compute the maximum value of the CHSH $S$--parameter \cite{CHSH} for the density matrix Eq. (\ref{1012}) \cite{Horodecky}. In Fig. 1, the range of parameters in which $S>2$ ---meaning that the CHSH inequality is violated--- is colored by light green, whereas the dark green part of the surfaces represents states in which $C>0$, but there is no violation of the CHSH inequality ($S\le2$). The thick gray line in Fig. 2 also marks the $S=2$ boundary. Below and to the left of this line $S>2$ for all $|\eta_1|^2$ values, but to its right $S$ may be smaller than 2 for some relative PSP orientations. Intriguingly, this non-locality boundary (thick gray line) nearly perfectly reproduces the boundary to the ESD-free region (the white area) if the scale on both axes is stretched by a factor of 1.5.

To conclude, we have carried out what we believe to be the first quantitative analysis of the decoherence of polarization entangled photons propagating in optical fibers. Our study shows that  PMD leads to entanglement sudden death in all but a well defined restricted set of realizations of fiber birefringence. In addition, we demonstrated how decoherence-free subspaces can be reached via non-local PMD compensation. The ultimate limits imposed by fiber birefringence to applications based on non-local properties of polarization entanglement were shown to be intriguingly related with the phenomenon of entanglement sudden death.

\end{document}

%\preprint{}

\title{Supplementary material for the manuscript entitled

``Sudden Death of Entanglement induced by Polarization Mode Dispersion''}
\author{Cristian Antonelli}%
\email{cristian.antonelli@univaq.it}
\affiliation{%
Dipartimento di Ingegneria Elettrica e dell'Informazione and CNISM, Universit\`a dell'Aquila, 67040 L'Aquila,
Italy
}%
\author{Mark Shtaif}
\affiliation{%
School of Electrical Engineering, Physical Electronics, Tel-Aviv University, Tel-Aviv 69978, Israel
}%
\author{Misha Brodsky}
\affiliation{%
AT\&T Labs, 200 Laurel Ave. S., Middletown, NJ 07748 USA
}%

\date{\today}

\begin{abstract}
In this supplement we derive analytical expressions for the purity, concurrence and S parameter of the Clauser-Horne-Shimony-Holt-type Bell's inequality, which characterize a pair of polarization entangled photons transmitted through two optical fibers with polarization mode dispersion.
\end{abstract}



\maketitle
\section{1. Introduction}
In our Letter \cite{Manuscript} we derive the density matrix describing the polarization state of two polarization-entangled photons transmitted over optical fibers with polarization mode dispersion (PMD). The density matrix of Eq. (7) in \cite{Manuscript} (we will further down refer to this equation as Eq. (M7)) is expressed in the frame of reference of the principal states of polarization (PSP), in which the derivation is greatly simplified and the final result appears quite compact. However, the analytical evaluation of some of the parameters characterizing the entanglement of the two-photon state is more convenient in a different frame of reference, in which the density matrix appears in the so-called $X$-shape \cite{X_YU_OC,X_YU_QIC}.

\iffalse In this supplement we provide analytical formulae for concurrence and CHSH $S$-parameter that, while having been used to carry out the analysis presented in the manuscript, have not been introduced there to comply with the four-page limit of PRL. \fi

In the first section of this supplement we study the purity of the two-photon state. In the following sections we provide analytical formulae for concurrence and CHSH $S$-parameter.

\section{2. Purity of the two-photon state}
The purity $\mathcal P$ of the two-photon state can be evaluated through straightforward algebra from Eq. (M7)
\bea \mathcal P &=&\frac{|\eta_1|^4+|\eta_2|^4}{2}+|\eta_1|^2|\eta_2|^2\left[|R_f(\tau_A)|^2+|R_f(\tau_B)|^2\right]\nonumber\\
&+& \frac{1}{2} \left[|\eta_1|^4|R_f(\tau_A-\tau_B)|^2 + |\eta_2|^4|R_f(\tau_A+\tau_B)|^2 \right].\eea
Obviously, the purity is equal to 1 when $\tau_A=\tau_B=0$, that is when there is no PMD at all. Yet interestingly, it remains equal to 1 when $|\eta_1|^2=1$ and the DGD values are equal, regardless of their magnitude. This property is a manifestation of the decoherence-free subspace concept. On the other hand, as the DGD values become larger than the inverse overlap bandwidth of Alice's and Bob's filters $B^{-1}$ (see \cite{Manuscript} for the definition of $B$), the purity approaches $|\eta_1|^4\left[ 1 + |R_f(\Delta \tau)|^2 \right] / 2 + |\eta_2|^4/2$, with $\Delta \tau = \tau_A - \tau_B$. Then, the minimum purity with respect to $\eta_1$ and $\eta_2$ is $\left[ 1 + |R_f(\Delta \tau)|^2 \right] / \left[ 2 + |R_f(\Delta \tau)|^2 \right] / 2$ and it ranges between $1/4$, for $\Delta \tau = 0$, and $1/3$, for $\Delta \tau \gg B^{-1}$.

\section{3. Expression for concurrence and $S$-parameter}
In this section we derive the expressions for the concurrence and the $S$ parameter of the CHSH version of Bell's inequality in terms of the parameters characterizing the PMD of the two fiber arms. This goal is most conveniently achieved by first expressing the density matrix in a basis where only the terms on its two diagonals are not zero, a form known as $X$-shape representation \cite{X_YU_OC,X_YU_QIC}. Subsequently the procedure requires defining the matrix $\mathbf S$ whose components are given by $s_{nm} = \mathrm{Trace}(\rho \sigma_n\otimes\sigma_m)$, where $\sigma_n$ ($n = 1,\,2,\,3$) are the Pauli matrices. Then the concurrence and the $S$ parameter are expressed in terms of the eigenvalues of $\mathbf{S}^T\mathbf{S}$, where $T$ denotes transposition. Eventually, the eigenvalues of $\mathbf{S}^T\mathbf{S}$ are related to the PMD parameters.

\subsection{3.1 The $X$-shape representation of the two-photon density matrix}
Since the individual photons are completely depolarized, the two-photon state belongs to the class of the Bell-diagonal states \cite{Horodecki96} and hence to that of the so-called $X$-states \cite{X_YU_OC,X_YU_QIC}. Namely, there exists a representation in which the density matrix takes the form of
\be \rho = \left[ \begin{array}{cccc}
                    \rho_{11} & 0 & 0 & \rho_{14} \\
                    0 & \rho_{22} & \rho_{23} & 0 \\
                    0 & \rho_{32} & \rho_{33} & 0 \\
                    \rho_{41} & 0 & 0 & \rho_{44}
                  \end{array}
 \right], \label{270}\ee
where all entries are real with $\rho_{11} = \rho_{44}$, $\rho_{22} = \rho_{33} = 1/2 - \rho_{11}$, $\rho_{14} = \rho_{41}$ and $\rho_{32} = \rho_{23}$.  In this representation concurrence is given by the formula \cite{X_YU_QIC}:
\be C = 2 \max\left\{0,|\rho_{23}| - \rho_{11}, |\rho_{14}| - \rho_{22} \right\}.  \label{273} \ee
The representation in which the density matrix is in the form of Eq. (\ref{270}) is constructed as follows. Using the coefficients $s_{nm} = \mathrm{Trace}(\rho \sigma_n\otimes\sigma_m)$ which form a $3\times 3$ real matrix $\mathbf S$, we find the eigenvectors of $\mathbf{S}^T\mathbf{S}$ and of $\mathbf{S}\mathbf{S}^T$, which could be viewed as vectors representing states of polarization in the Stokes space \cite{Gordon&Kogelnik}. One then chooses the two orthogonal Jones vectors that correspond to one of the eigenvectors of $\mathbf{S}^T\mathbf{S}$ to span the subspace corresponding to the first photon. Similarly an orthogonal pair of Jones vectors corresponding to one of the eigenvectors of $\mathbf{S}\mathbf{S}^T$ is chosen to span the subspace corresponding to the second photon. The overall space, which is the tensor product of the two subspaces, is then spanned by the two pairs of Jones vectors that were selected \cite{Horodecki96}. In this process, the eigenvector of $\mathbf{S}^T\mathbf{S}$ and the eigenvector of $\mathbf{S}\mathbf{S}^T$ that the desired representation is based upon, are selected arbitrarily. Moreover, it is irrelevant which of the two pairs of Jones vectors is assumed to represent a specific photon's subspace, namely the two pairs can always be interchanged without compromising the process.

In this representation one can verify, using Eq. (\ref{270}) along with the definition of the $s_{nm}$ elements, that $\mathbf{S}$ is diagonal and therefore
\be \mathbf S^T \mathbf S = \frac 1 4 \left[ \begin{array}{ccc}
                                     (\rho_{14}+\rho_{23})^2 & 0 & 0 \\
                                     0 & (\rho_{14}-\rho_{23})^2 & 0 \\
                                     0 & 0 & (\rho_{11}-\rho_{22})^2
                                   \end{array}
 \right] \label{272}\ee
This form will be later used for relating the concurrence and the $S$ parameter to the quantities characterizing PMD through the eigenvalues of $\mathbf S^T \mathbf S$.

\subsection{3.2 Expressing concurrence in terms of the eigenvalues of $\mathbf{S}^T\mathbf{S}$ }
The aim of this section is to demonstrate the following expression for concurrence in terms of the eigenvalues of $\mathbf{S}^T\mathbf{S}$:
\be C = \frac 1 2 \max\left\{  0 , \lambda_1^{1/2} + \lambda_2^{1/2} + \lambda_3^{1/2} -1  \right\}, \label{271}\ee
where we denote these eigenvalues as $\lambda_1$, $\lambda_2$ and $\lambda_3$. For the calculation that follows, it is convenient to assume that the eigenvalues are sorted such that $\lambda_1\ge\lambda_2\ge\lambda_3$.

Our first goal is to express the Bell-diagonal density matrix in terms of $\lambda_1$, $\lambda_2$ and $\lambda_3$. Let us consider a basis for subspace $A$ such that the corresponding Stokes \textit{unit} vector $\hat b_A$ is an eigenvector of $\mathbf{S}^T\mathbf{S}$ belonging to the largest eigenvalue, namely $\mathbf{S}^T\mathbf{S} \hat b_A = \lambda_1 \hat b_A$.  Consider also a basis for subspace $B$ such that the corresponding Stokes unit vector $\hat b_B$ is an eigenvector of $\mathbf{S}\mathbf{S}^T$ defined though $\hat b_B = \mathbf S \hat b_A / ||  \mathbf S \hat b_A  ||$, where $||\cdot ||$ denotes vector length.

We denote the correlation between the measurements of coincident photons along the polarizations defined by the Stokes vectors $\hat b_A$ and $\hat b_B$ as $E(\hat b_A,\hat b_B) = P_c - (1-P_c)$, where $P_c = \rho_{11} + \rho_{44} = 2 \rho_{11}$ is the probability of a coincident detection of photons $A$ and $B$ passing through polarizer beam splitters aligned to $\hat b_A$ and $\hat b_B$. From Eq. (7) of \cite{Horodecki96}, $E(\hat b_A,\hat b_B) = \hat b_A \cdot \mathbf S \hat b_B = \lambda_1^{1/2}$, it follows that $\rho_{11} = \rho_{44} = (1 + \lambda_1^{1/2})/4$. Also from the condition of unit trace of the density matrix, $\rho_{22} = \rho_{33} = (1 - \lambda_1^{1/2})/4$. This yields $\lambda_1$ as the third diagonal element of $\mathbf{S}^T\mathbf{S}$.

The remaining off-diagonal terms of the density matrix follow from equating the first and the second diagonal elements of $\mathbf{S}^T\mathbf{S}$ to $\lambda_2$ and $\lambda_3$ (in some arbitrarily assumed order). This yields two possible solutions: either $|\rho_{23}| = |\rho_{32}| = \lambda_2^{1/2} + \lambda_3^{1/2}$ and $|\rho_{14}| = |\rho_{41}| = \lambda_2^{1/2} - \lambda_3^{1/2}$, or   $|\rho_{23}| = |\rho_{32}| = \lambda_2^{1/2} - \lambda_3^{1/2}$ and $|\rho_{14}| = |\rho_{41}| = \lambda_2^{1/2} + \lambda_3^{1/2}$. However, only one of them is compatible with nonzero values of concurrence. In fact, using Eq. (\ref{273}) for the choice $|\rho_{23}| = |\rho_{32}| = \lambda_2^{1/2} + \lambda_3^{1/2}$ and  $|\rho_{14}| = |\rho_{41}| = \lambda_2^{1/2} - \lambda_3^{1/2}$ yields $C =  \max\left\{0, -\lambda_1^{1/2} + \lambda_2^{1/2} + \lambda_3^{1/2} - 1 , \lambda_1^{1/2} + \lambda_2^{1/2} - \lambda_3^{1/2} - 1  \right\} / 2 = 0$. Indeed, since all eigenvalues are less than unity and $\lambda_1$ is the largest, the following inequality is true, $-\lambda_1^{1/2} + \lambda_2^{1/2} + \lambda_3^{1/2} - 1 \le 0$. On the other hand, the positivity of $\rho$ implies that $\lambda_2^{1/2} - \lambda_3^{1/2} \le 1-\lambda_1^{1/2}$, which yields $\lambda_1^{1/2} + \lambda_2^{1/2} - \lambda_3^{1/2} - 1 \le 0$. Hence the correct solution is $|\rho_{23}| = |\rho_{32}| = \lambda_2^{1/2} - \lambda_3^{1/2}$ and $|\rho_{14}| = |\rho_{41}| = \lambda_2^{1/2} + \lambda_3^{1/2}$ and the density matrix reads as:
\be \rho = \frac 1 4 \left[ \begin{array}{cccc}
                    1+\lambda_1^{1/2} & 0 & 0 & \zeta(\lambda_2^{1/2}+\lambda_3^{1/2}) \\
                    0 & 1-\lambda_1^{1/2} & \xi(\lambda_2^{1/2}-\lambda_3^{1/2}) & 0 \\
                    0 & \xi(\lambda_2^{1/2}-\lambda_3^{1/2}) & 1-\lambda_1^{1/2} & 0 \\
                    \zeta(\lambda_2^{1/2}+\lambda_3^{1/2}) & 0 & 0 & 1+\lambda_1^{1/2}
                  \end{array}
 \right]. \label{280}\ee
where $\zeta$ and $\xi$ can arbitrarily be set to $\pm1$. This indetermination reflects the phase ambiguity in the conversion from Stokes to Jones representation of a polarization state \cite{Gordon&Kogelnik}.

Finally, applying the simplified concurrence formula Eq. (\ref{273}) to the density matrix in Eq. (\ref{280}) yields the expression for concurrence given in Eq. (\ref{271}).

\subsection{3.3. Expressions for the eigenvalues of $\mathbf{S}^T\mathbf{S}$ in terms of the PMD parameters}
The analytical derivation of the eigenvalues $\lambda_i$ requires to solve a third order algebraic equation. This task is in general quite cumbersome. However, for the density matrix in the shape of Eq. (M7) this is possible under two conditions that make the matrix real. First, we make a realistic assumption on the symmetry of Alice's and Bob's filters, $|H_{A}(\omega)H_{B}(-\omega)| = |H_{A}(-\omega)H_{B}(\omega)|$. Under this assumption $R_f(\tau)$ is real. Secondly, we perform a change of basis. In particular, we use a basis $(|\uu{\tilde s}_A \rip,|\uu{\tilde s}'_A \rip)$ for the subspace $A$ where $|\uu{\tilde s}_A \rip = e^{i\varphi_A/2} |\uu{s}_A \rip$  and $|\uu{\tilde s}'_A \rip = e^{-i\varphi_A/2} |\uu{s}'_A \rip$ with $\varphi_A = \alpha + \mathrm{Im}[\log(\eta_1\eta_2)]$. Similarly, we use a basis $(|\uu{\tilde s}_B \rip,|\uu{\tilde s}'_B \rip)$ for the subspace $B$ where $|\uu{\tilde s}_B \rip = e^{i\varphi_B/2} |\uu{s}_B \rip$ and $|\uu{\tilde s}'_B \rip = e^{-i\varphi_B/2} |\uu{s}'_B \rip$ with $\varphi_B = \mathrm{Im}[\log(\eta_1\eta_2)]$. In this representation the density matrix of Eq. (M7) can be simplified by setting $\alpha = 0$ and by replacing $\eta_1$ and $\eta_2$ with their absolute values $|\eta_1|$ and $|\eta_2|$, thus making the matrix real. In this representation the matrix $\mathbf S^T \mathbf S$ appears as:
\be \mathbf S^T \mathbf S = \left[ \begin{array}{ccc}
                                     m_{11} & 0 & m_{13} \\
                                     0 & m_{22} & 0 \\
                                     m_{13} & 0 & m_{33}
                                   \end{array}
 \right] \label{300}\ee
with
\bea m_{11} &=& \left[|\eta_1|^2 R_f(\tau_A- \tau_B)-|\eta_2|^2 R_f(\tau_A + \tau_B)\right]^2+4|\eta_1|^2 |\eta_2|^2 R_f^2(\tau_B) \\ m_{22} &=& \left[|\eta_1|^2 R_f(\tau_A- \tau_B)+|\eta_2|^2 R_f(\tau_A+ \tau_B)\right]^2 \\ m_{33} &=& 4 |\eta_1|^2 |\eta_2|^2 R_f^2(\tau_A)+(2|\eta_1|^2-1)^2 \\ m_{13} &=& -2\left[|\eta_1|^2 R_f(\tau_A-\tau_B)-|\eta_2|^2 R_f(\tau_A + \tau_B) \right] |\eta_1| |\eta_2| R_f(\tau_A) \nonumber \\ &+& 2|\eta_1\eta_2| R_f(\tau_B) (2|\eta_1|^2-1)  \label{310}\eea
The eigenvalues of $\mathbf S^T \mathbf S$ can be evaluated with the following result
\bea \ell_1 &=& (m_{11} + m_{33})/2 + \left[(m_{11} + m_{33})^2/4 - (m_{11} m_{33}-m_{13}^2)\right]^{1/2} \\ \ell_2 &=& m_{22} \\ \ell_3 &=& (m_{11} + m_{33})/2 - \left[(m_{11} + m_{33})^2/4 - (m_{11} m_{33}-m_{13}^2)\right]^{1/2},  \label{320} \eea
where $\ell_1$, $\ell_2$ and $\ell_3$ are the unsorted eigenvalues of $\mathbf S^T \mathbf S$. The concurrence is then given by
\be C = \frac 1 2 \max\left\{  0 , \ell_1^{1/2} + \ell_2^{1/2} + \ell_3^{1/2} -1  \right\}. \label{330} \ee
The largest violation of CHSH inequality is given by $S = 2\sqrt{\lambda_1 + \lambda_2}$, where $\lambda_1$ and $ \lambda_2$ are the two largest eigenvalues of  $\mathbf S^T \mathbf S$ \cite{Horodecki}. From Eqs. (12) and (14) it follows that $\ell_1 > \ell_3$, and therefore
\be S = 2\sqrt{\ell_1+ \max\left\{ \ell_2 , \ell_3 \right\} }. \label{340} \ee
Equations (\ref{330}) and (\ref{340}) are the main result of this supplement. They give the expressions for largest violation of Bell's inequality and concurrence as a function of the instantaneous PMD values.
\subsection{3.4 Entanglement sudden death threshold for large differential group delays}
We conclude this supplement by evaluating the concurrence of the output two-photon state in a particular case of interest, that is when the two DGDs are much larger than the inverse photon bandwidth. In this regime one can derive the following expression
\bea C=\max\Big\{0,\big[1+|R_f(\Delta\tau)|\big]|\eta_1|^2-1\Big\}\label{Conc},\eea
where $\Delta\tau$ is the difference between the two DGDs. This results can be used to characterize a threshold seen in Fig. 1a of the Letter. In that figure it appears that for two equal DGDs there is a threshold value of $|\eta_1|^2=0.5$, above which ESD does not occur for increasing DGDs. In fact, such a threshold exists as long as the difference between two DGDs is finite. Using the above asymptotic expression for concurrence this threshold can be estimated as $|\eta_{1,\mathrm{th}}|^2=[1+|R_f(\Delta\tau)|\big]^{-1}$.